
\input harvmac.tex
\def\Title#1#2{\rightline{#1}\ifx\answ\bigans\nopagenumbers\pageno0\vskip1in
\else\pageno1\vskip.8in\fi \centerline{\titlefont #2}\vskip .5in}

\def\thetaf#1{\theta_#1}

\def\p {\partial}
\def\CN {{\cal N}}

\def\inbar{\,\vrule height1.5ex width.4pt depth0pt}
\def\IB{\relax{\rm I\kern-.18em B}}
\def\IC{\relax\hbox{$\inbar\kern-.3em{\rm C}$}}
\def\IP{\relax{\rm I\kern-.18em P}}
\def\IR{\relax{\rm I\kern-.18em R}}
\font\cmss=cmss10 \font\cmsss=cmss10 at 7pt
\def\IZ{\relax\ifmmode\mathchoice
{\hbox{\cmss Z\kern-.4em Z}}{\hbox{\cmss Z\kern-.4em Z}}
{\lower.9pt\hbox{\cmsss Z\kern-.4em Z}}
{\lower1.2pt\hbox{\cmsss Z\kern-.4em Z}}\else{\cmss Z\kern-.4em Z}\fi}

\def\tr{{\rm tr~}}
\def\half{{1\over2}} 

\Title{\vbox{\baselineskip12pt\hbox{LPTENS-93/20}\hbox{RU-93-17}}}%
{\vbox{\centerline{Large $N$ Phase Transition in Continuum QCD$_2$}}}

\centerline{Michael R. Douglas$^1$ and Vladimir A. Kazakov$^2$}
\bigskip
\centerline{Laboratoire de Physique Theorique}
\centerline{D\'epartment de Physique de l'Ecole Normale Sup\'erieure}
\centerline{24 rue Lhomond, 75231 Paris Cedex 05, France}
\bigskip
\centerline{Department of Physics and Astronomy}
\centerline{Rutgers University, Piscataway, NJ 08854}
\bigskip

We compute the exact partition function for pure
continuous Yang-Mills theory on the
two-sphere in the large $N$ limit, and find that it exhibits a large $N$
third order phase transition with respect to the area $A$ of the sphere.
The weak coupling (small A) partition function
is trivial, while in the strong coupling phase (large A) it is
expressed in terms of elliptic integrals.
We expand the strong coupling result in a double power series in
$e^{-g^2 A}$ and $g^2 A$ and show that the terms are the weighted sums of
branched coverings proposed by Gross and Taylor.
The Wilson loop in the weak coupling phase does not show the simple area
law.

We discuss some consequences for higher dimensions.

\footnote{}{$^1$ (mrd@physics.rutgers.edu)}
\footnote{}{$^2$ (kazakov@physique.ens.fr)}

\Date{May 11, 1993}
\nref\thooft{G.'t Hooft, Nucl.Phys B72 (1974) 461}
\nref\wilson{ K.Wilson, Phys.Rev. D8 (1974) 2445 }
It is an old and famous idea, that large $N$ QCD is equivalent to a string
theory. \refs{\thooft,\wilson}
Many approaches have been tried to make this precise.
One of the most promising is to interpret the diagrams of the strong coupling
expansion of the Euclidean lattice theory as string world sheets.
This combines the great advantage of the finite $N$ strong coupling expansion,
that confinement is already present at leading order,
with the great advantage of large $N$, that we have a free string theory at
leading order.
This expansion was proposed for
any dimension in
\ref\katwo{V.A.Kazakov, Sov.Phys.JETP, 58(6) (December 1983).
1986, Phys.Lett. 128B (1983) 316.}
\nref\zu{J.-B.Zuber and K.H. O'Brien, Nucl. Phys. B253 (1985) 621-634.}
\nref\koone{I.K.Kostov, Nucl.Phys.B265 (1986) 86, Phys.Lett. 138B
(1984) 191}
and elaborated in \refs{\zu,\koone}.

Early hopes for the use of the strong coupling
expansion were dashed by the discovery of Gross and Witten
and of Wadia
\nref\GW{D.J. Gross and E. Witten, Phys.Rev.D21 (1980) 446-453.}
\nref\wadia{S. R. Wadia, Phys.Lett.93B (1980) 403.}
\refs{\GW,\wadia},
that even the simplest integrals involving the Wilson action, such as
\eqn\ww{\int dU~e^{N/g^2 \tr (U+U^{+})} ~\tr U}
(in two-dimensional Yang-Mills theory, this is a Wilson loop enclosing a single
plaquette of area $1$), were non-analytic in the coupling constant (see
\ref\gb{E.Brezin and D.Gross, Phys.Lett.B97 (1980) 120} for the
complete treatment of the problem).

Such behavior is possible because we have $O(N)$ degrees of freedom, and these
integrals are dominated by a saddle point.  Just as for the infinite volume
limit in statistical mechanics, several saddle points may exist, and the large
$N$ limit picks the one of lowest action for a given value of the coupling.
The consequence in higher dimensions was that integration formulas used in
developing the strong coupling expansion were only valid down to a critical
$g_c^2$.  The expansion had no validity in the physical regime of weak bare
coupling.

It was not then clear whether this transition is a fundamental barrier to
combining strong coupling and large $N$ or just a reason not to use the Wilson
action in this context.
Although we now know much more about large $N$ phase transitions, after having
studied them in depth for their application to two-dimensional gravity and
string theory (this one was studied in
\ref\neub{H. Neuberger, Nucl. Phys. B340 (1990) 703; V. Periwal and D. Shevitz,
Phys. Rev. Lett. 64 (1990) 1326; C. Crnkovic, M. Douglas and G. Moore, Nucl.
Phys. B360 (1991) 507-523.}),
this point is still not clear.
It is therefore interesting to modify the action and see  if this modifies or
eliminates the transition.  The only obvious constraint is the very weak one
that for small lattice spacing, the $\tr F^2$ term should be present in the
action, with higher dimension terms not unnaturally enhanced.

The point was made by Migdal
\ref\mig{A.A.Migdal, Sov. Phys. JETP 42 (1975) 413; 743.}
that in two dimensions, one could exactly integrate out a link common to two
plaquettes, and compute an exact renormalization group transformation.  The
Wilson action is not a fixed point of this transformation; it and generic
nearby actions flow to the ``heat kernel action,'' more properly described as a
Boltzmann weight for a plaquette with holonomy $U$:
\eqn\hk{Z_{hk}(U;g^2 A) = \sum_R \dim R e^{-g^2 A C_2(R)/2N} \chi_R(U)}
which is the heat kernel on the group manifold $G(U,1;\Delta t=g^2 A/2N)$.
This was also proposed in
\ref\meno{P. Menotti and E. Onofri, Nucl.Phys. B190 (1981) 288-300.}
for its other nice properties, among them being that it gives exact equivalence
between Euclidean and Hamiltonian lattice formulations, and that
using the heat kernel action, the above Wilson loop expectation value suffers
no large $N$ transition.
{}From this one might conjecture that the transition was a lattice artifact,
and
that using the heat kernel action gives us continuum answers with no
transition.
(Of course this action, used for the plaquettes of a regular lattice, is not an
RG fixed point in $D>2$, so even if it worked for $D=2$ the conjecture would
not be proven.)

Already in two dimensions the continuum QCD observables, Wilson loop averages:
\eqn\loop{W(C_1,C_2,...,C_k) = <\prod_{j=1}^k \tr P\exp[i \oint_{C_j} d x_\mu
A_\mu(x)] >}
(where $C_j$ are arbitrarily intersecting and selfintersecting
contours on the infinite plane), show a great deal of structure.
These were first computed by
\ref\kkone{V.A.Kazakov and I.K.Kostov, Nucl.Phys.B176
(1980) 199} for the $U(N)$ gauge group with $N \rightarrow \infty$, and
generalized to any $N$ in \ref\ka1{V.A.Kazakov, Nucl.Phys.B179 (1981) 283}
and to the lattice version of the theory
\ref\kk2{V.A.Kazakov and I.K.Kostov, Phys.Lett.105B (1981)453}
\nref\bra{N.Bralic, Phys.Rev. D22 (1980) 3090.}
\nref\rossi{P.Rossi, Ann.Phys 132 (1981) 463.}
(see also \refs{\bra,\rossi}). This
calculation was based on the renormalized
two-dimensional version of the Makeenko-Migdal loop equations
\ref\mm{Yu.Makeenko and A.A.Migdal,
Phys.Lett.88B (1979) 135} established in \kkone~ and (in modern
language) on the invariance of the results under area preserving
diffeomorphisms.

    The most striking feature of the Wilson averages in the limit $N
\rightarrow \infty$, observed in \kkone~ was their ``stringy" character:
each of them was shown to be a sum over all possible surfaces of the
minimal area (without folds) spanned on the contour (``soap films"), of
an exponent of minus area of the surface (area law) times some polynomial
of the areas of domains forming this surface.
The geometrical
interpretation of the polynomials was not known at the time.
It became clearer from the paper of I.~Kostov \koone~(see Appendix A
there)
 where it was demonstrated
that they come from the statistics of surfaces (coverings with boundary)
having branch points and cuts, connecting various sheets of a surface.
For recent developments in this direction see
\ref\newkos{I.K.Kostov, preprint S.Ph.T.-93-060 (1993, May)}.

\nref\mrd{M. R. Douglas, preprint RU-93-13, hep-th/9303159.}
Another nontrivial quantity is the partition function
on a two dimensional compact manifold of area
$A$ and genus G. A formula in terms of the sum over
representations of the gauge group,
\eqn\qcd{Z_G(g^2 A) = \sum_R (\dim R)^{2-2G} e^{-g^2 A C_2(R)/2N}.}
was found in
\ref\rus{B.Rusakov, Mod.Phys.Lett. A5 (1990) 693.}.
The sum for the case of the group $U(N)$, say, goes over all Young
tableaux characterized by the components of the highest weight
$\{n_1,n_2,...,n_N\}$ which are the integers obeying the inequality:
\eqn\ineq{\infty \ge n_1 \ge ... \ge n_N \ge -\infty}
So we see that it is still a complicated multiple sum which takes some
effort to calculate in particular cases.

\nref\gr{D. Gross, Princeton preprint PUPT-1356, hep-th/9212149}
\nref\gt{D. Gross and W. Taylor, preprints PUPT-1376 hep-th/9301068 and
PUPT-1382 hep-th/9303046.}
A nice interpretation of this partition functions in terms of minimal
coverings, similar to those of \kkone~and \koone~, was given by Gross and
Taylor. \refs{\gr,\gt}~
It was noticed that the partition function can be written in terms of a
sum over minimal coverings of a manifold of a genus G and area A, where
different sheets of a covering are glued together with elements
called branch points, tubes and omega-points. Using an inequality
known by mathematicians, about the possibility of minimal covering of
the manifold of a genus G by a surface of a genus g, Gross found that many
terms of the $1/N$ topological expansion in the free energy $F(A,N)=
{1 \over N^2} \log Z(A,N)$ are equal to zero. So, for $G=1$ he found
that the $O(N^2)$ (spherical world-sheet) contribution to $F(A,N)$ is zero.
However, for the next order (torus
topology of coverings) the sum over minimal coverings is infinite and
can be given in terms of the Dedekind function:
\eqn\torus{F = -2\log \eta(iA/4\pi)
 = -{A\over24}-2 \sum_{n\ge 1}{\log (1-e^{-n A/2})}}
(the constant is of course a choice of ground state energy.)

      For the spherical topology of the space $G=0$, the result is more
complicated to obtain. Its geometrical interpretation contains all of the
additional elements mentioned above, and a
purely geometrical derivation looks tricky.

     In this paper we present the explicit result for the
leading order (planar) contribution to the free energy of the
Yang-Mills theory on the two-dimensional sphere.
For large area of the sphere (compared to $1/g^2$) it nicely fits the
interpretation in terms of minimal coverings, down to the phase transition
point $g^2 A_{crit} =\pi^2$, where the
sum over coverings is divergent.
In the phase of small $g^2 A$ the result
is trivial.

The partition function is
\eqn\parti{  Z_{G=0}(A,N=\infty ) = \exp[N^2 F(A)] =
  \sum_R (\dim R)^2 e^{- {A \over 2 N} C_2(R)}}
where we measure the area $A$ in units of $1/g^2$, and for the
group $U(N)$ we have
\eqn\dimcasi{\eqalign{C_2(R) &= \sum_{i=1}^{N} n_i ( n_i - 2 i +N+1 )\hfill\cr
\dim R &= \prod_{i>j} (1- {n_i - n_j \over i-j } ) }}
and the sum over the representations R has to be understood as a
multiple sum over $N$ integer variables ${n_1,...,n_N}$ obeying the
inequality \ineq.
\footnote{$\dagger$}{Strictly speaking these are only representations whose
$U(1)$ charge is correlated with the charge under the center of $SU(N)$, in
other words of $SU(N)\times U(1)/\IZ_N$.  Since a single representation will
dominate in the final answer, it is also correct for $U(N)$.}
Now, in the large $N$ limit, nothing prevents us from using
continuum variables:
\eqn\cont{n(x)={n_i \over N}, \ \ \ \ \ x={i \over N},    }
obeying now the inequality:
\eqn\inecon{n(x) \ge n(y) , \ \ \ \ {\rm if} \ \ \ \  x \le y.    }
It is convenient to change the variable to
\eqn\change{h(x)=-n(x)+x-1/2} and to write formally the following
functional integral representation for the partition function:
\eqn\func{ Z_0(A) = \int D h(x) \exp -N^2 S_{eff}[h(x)]   }
where
\eqn\seff{ S_{eff}[h(x)] = -\int_0^1 dx \int_0^1 dy \log |h(x)-h(y)| +
   {A \over 2} \int_0^1 dx h^2(x) - A/24   }
and $h(x)$ obeys, according to \ineq~ and \change~ , the inequality
\eqn\cond{  {h(x) - h(y) \over x-y } \ge 1.  }
One of the main observations of this paper is the need to respect
this condition in calculating \func, which will lead to nontrivial
consequences such as a new large N phase transition and the
existence of the large area (strong coupling) phase.
If we introduce the density of the boxes in the Young tableau in terms
of the variable $h$
\eqn\den{ u(h) = { \p x(h) \over \p h } }
with the normalization
\eqn\norm{ \int dh~ u(h) =1 }
the condition \cond~ can be simply rewritten as
\eqn\cons{ u(h) \le 1 , \ \ \ \ {\rm for \ \ any} \ \ h  }

Since we have a large parameter $N^2$ in front of the $S_{eff}$ in
\func~ we can try to apply the saddle point approximation, which means
that we have to solve the equation on $h(x)$
\eqn\saddle{ {\delta S_{eff}[h(x)] \over \delta h(x) } =0  }
Let us ignore for a moment the constraint \cons~. This will lead us
immediately to the integral equation
\eqn\semi{{A\over 2} h =  P\int {ds u(s) \over h-s }  }
which is precisely the same as for the distribution of the eigenvalues
in the hermitean gaussian matrix model. The solution is the well-known
semi-circle law of Wigner:
\eqn\wi{u(h) = {A \over 2 \pi } \sqrt{ {4 \over A} - h^2 } }

{}From here we obtain immediately for the derivative of the free energy
with respect to the area $A$
\eqn\fweak{  F'(A) = -{ \p S_{eff}[h^*]  \over \p A } =
< { \tr \over 2N } h^2 > -{1\over 24} = {{1\over 2A}  - {1\over 24}  }  }
or
\eqn\ff{ F(A) = {1\over 24}A - \half \log A   }
This result is already clear from \func~ where the $\log A$ term can be
obtained by simple re-scaling of the continuous field $h$ by
$\sqrt{A}$, after which the gaussian integral will not depend on $A$.
This result was obtained by the same reasoning
in \ref\rustwo{B.Rusakov, Phys.Lett. B303 (1993) 95.}, where
the inequality \cons~ was ignored, and hence, the existence of the
second, strong coupling phase was not noticed.

Let us explain this result from another point of view.
Clearly from \hk, we have
\eqn\group{Z_{G=0}(A) = Z_{hk}(U=1;\ A) = G(U=1,~U'=1;\ A/2N).}
For small $A$, we really are interested in the small time asymptotics of the
heat kernel on the group manifold.
These are well known for arbitrary group manifold $H$:
\eqn\groupasy{Z_{G=0}(A) \sim
\left({N\over 2\pi A}\right)^{\dim H/2} + O(e^{-2\pi^2 N/A})}
where the correction is the sum of $e^{-S}$ over all closed orbits on $H$.

We see the origin of the term $-\half \log A$, and that the corrections are
suppressed as $e^{-N/A}$, hence vanishing in the large $N$ limit.

Remembering now that we have the constraint \cons,
we note that the trivial solution \wi~ is
possible only for small areas $A$:
\eqn\acrit{ A \le A_{crit} = \pi^2   }

What happens for $A > A_{crit}$? We still have to solve the saddle
point equation \saddle~ , but in the presence of the boundary
conditions \ineq~, or, which is the same, \cons~. The only thing which
may happen is that a finite fraction of the highest weight components
${n_1,...n_N}$  will condense at the boundary of the inequality, namely
\eqn\gap{ n_{k+1}=n_{k+2}=...=n_{N-k} =0}
whereas all others are non-zero (see fig.1b and compare it to
fig.1a, where a typical Young tableau for the weak coupling phase
is presented).

On the language of the density for the continuous variable $h$ it
means that
\eqn\denst{\eqalign{
 u(h) &= 1, \ \ \ \ \ for \ \ -b \le h \le b    \cr
      &= \tilde{u}(x) \ \ \ \ \ elsewhere  }}
where $b=1/2-k/N$ is finite in the large $N$ limit and
 $\tilde{u}$ is some nontrivial function to be found later.

In fig. 2 the Young tableau is presented in both phases for the
variable $h(x)$. We are trying to find here a so called two-cut
solution, similar to those observed in the hermitean matrix models
with double well potentials
\ref\twoc{E. Br\'ezin, unpublished; G.M. Cicuta, L. Molinari and E. Montaldi,
Mod. Phys. Lett. A 1 (1986) 125.}.

Let us substitute the ansatz \denst~ into \semi~ . We get the
following equation on $\tilde{u}(h)$:
\eqn\veff{A/2 h - \log {h-b \over h+b} =
 P\int {ds \tilde{u}(s) \over h-s }  }

As we see now, the condensate of the zero highest weight components
induces an extra logarithmic term in the equation. In the language of
the equivalent hermitean one matrix model it would mean that we have the
effective matrix potential whose derivative is the l.h.s. of \veff~.
This potential clearly has two wells separated from each other by the
cut of logarithm. The eigenvalues fill these wells always to the top.
All the eigenvalues which spill over the top (say, by increasing $A$)
form the condensate.

We introduce, as usual, the function of the complex variable $h$
\eqn\compl{ f(h) = \int ds {\tilde{u}(s)  \over  h-s}  }
whose imaginary part is  $\pi \tilde{u}(h)$.
The solution of the corresponding two-cut Cauchy problem
\ref\gakh{D.Gakhov, Boundary problems, Russian edition ``Nauka" (1975);
A.C.Pipkin, A Course on Integral Equations, Springer-Verlag, 1991.}
is given as a contour integral
\eqn\fu{f(h) =  - { 1 \over 2 \pi i } \sqrt{(a^2-h^2)(b^2-h^2)}
\oint ds { \half A s - \log {h-b \over h+b} \over  (h - s)
\sqrt{(a^2-s^2)(b^2-s^2)}  }  }
where the contour of integration encircles the cuts of the square root
but leaves aside the singularities of the nominator and of the pole at
$s=h$. The limits $a$ and $b$ have to be found later from the
condition of the correct behaviour of $f(h)$ for $h \rightarrow
\infty$.

Now, by inflating the contour we catch, instead of the cuts of the
square root, the singularity at the pole and the cut of the logarithm.
This gives
\eqn\jump{ f(h) =  h {A \over 2} - \log {h-b \over h+b} +
 \sqrt{(a^2-h^2)(b^2-h^2)}
\int_{-b}^b ds {1 \over  (h - s)
\sqrt{(a^2-s^2)(b^2-s^2)}  }  }

Let us note that the imaginary part of the logarithm in the r.h.s. of
\jump~ is exactly equal to the minus $\pi$ times density of condesate of
 $h's$ in the interval $[-b,b]$. So it is clear that the last term in
\jump~ represents the full function $u(h)$ defined by \denst~. The
latter is expressible in terms of the complete elliptic integral of
the third kind $\Pi [\theta, x]$ :
\eqn\denfin{u(h) = {1 \over \pi}{b-a \over b+a } \sqrt{{(a+h)(b+h)
\over
(a-h)(b-h)}} \Pi [{2b \over a+b}{h-a \over h+b},{2\sqrt{ab} \over
a+b}]    }

In order to find the parameters $a$,$b$ and the whole free energy, it
is better to use the asymptotics for the large $h$ of \jump~
\eqn\assa{\eqalign{  f(h) =
& h \big( {A \over 2} - \int_{-b}^b ds
{1 \over \sqrt{(a^2-s^2)(b^2-s^2)} } \big)   \cr
& +h^{-1} \big( 2b +  \int_{-b}^b ds {s^2- {a^2+b^2 \over 2}
 \over \sqrt{(a^2-s^2)(b^2-s^2)}} \big)  +                     \cr
& + h^{-3} \big(2b +  \int_{-b}^b ds {s^4- {a^2+b^2 \over 2}
s^2-{(a^2-b^2)^2
\over 8}
 \over \sqrt{(a^2-s^2)(b^2-s^2)}} \big)  +  O(h^{-5})   \cr   }}
and compare it with that which follows from the definition \compl~:
\eqn\assb{ f(h) = 0 \cdot h - (1-2b) h^{-1} - (F'(A) + 1/24 ) h^{-3}
 + O(h^{-5})    }
This comparison gives (the elliptic integrals are in the appendix) the
following results:
The first derivative of the  free energy in the strong coupling phase,
expressed in terms of elliptic integrals with modulus $k=b/a$, is
\eqn\fprime{F'(A) = {1\over 6}a^2 - {1\over 12}a^2 k'^2 - {1\over 24}
+ {1\over 96} a^4 k'^4 A}
where the modulus is related to the area by
\eqn\area{{1\over 4}A = (2E - k'^2 K) K,}
the complementary modulus $k'^2=1-k^2$, and
\eqn\limit{  a = 4 K/A.}

This solution represents the strong coupling phase of our theory,
namely for the area of sphere $A \ge \pi^2$. It is easy to check that
at the point of transition $A_{crit} = \pi^2$ the two solutions coincide
completely, even for the distribution $u(h)$ of boxes in the Young
tableau. Let us calculate the order of this transition.

Series expansions become easier in terms of theta constants for a torus of
complex modulus $\tau$.
The equation \area~ becomes (the relevant identities are in the appendix)
\eqn\areax{\eqalign{
A &= 8EK-4k'^2 K^2\hfill\cr
&= {\pi^2\over 3}
(\thetaf{2}^4(0|\tau) + \thetaf{3}^4(0|\tau) + 2 E_2(\tau))\hfill\cr
&= \pi^2(1 + 8 q - 8 q^2 + 32 q^3 + \ldots)}}
where $q=e^{i\pi\tau}$, and the critical point is the limit $\tau\rightarrow
i\infty$.

Now
\eqn\ffd{\eqalign{F'_{\rm strong}(A) - F'_{\rm weak}(A) &=
{\pi^2\over 3A^2}(\thetaf{2}^4(0|\tau) + \thetaf{3}^4(0|\tau))
+ {\pi^4\over 6A^3}\theta_4^8(0|\tau) - {1\over 2A} -{1\over 24}\hfill\cr
&={1\over 2A} - {2\pi^2\over 3 A^2}E_2(\tau) + {\pi^4\over 6A^3}
\theta_4^8(0|\tau)\hfill\cr
&\rightarrow 0 ~~~{\rm as}~~~ \tau\rightarrow i\infty}}
so we see that the transition is higher order.
Inverting \areax~ and substituting,
\eqn\fdx{\eqalign{F'_{\rm strong}(A) - F'_{\rm weak}(A) =&
{1\over 2A} - {2\pi^2\over 3 A^2}
(1-{3\over 8}\left({A-A_c\over\pi^2}\right)^2-{3\over
32}\left({A-A_c\over\pi^2}\right)^3+\ldots)\hfill\cr
&+ {\pi^4\over 6A^3} (1-2\left({A-A_c\over\pi^2}\right)+{3\over
2}\left({A-A_c\over\pi^2}\right)^2+\ldots)\hfill\cr
=&{1\over \pi^2} \left({A-A_c\over\pi^2}\right)^2 + \ldots}}

Thus the phase transition is of the third order,
like the well known Gross-Witten-Wadia phase transition for the lattice two
dimensional multicolour gauge theory. However, in spite of some
similarities of these two transitions, the one found in this paper
happens already in the continuum version of the theory, so we cannot
say that it is a lattice artifact.

The transition also bears some similarity with the
Berezinski-Kosterlitz-Thouless transition of condensation of vortices
on the world sheet of one-dimensional string theory compactified on a
circle \ref\klegr{D.Gross and I.Klebanov, Nucl.Phys.B354 (1990) 459   }.
In the language of the corresponding matrix quantum mechanics
the point of the phase transition also corresponds there to the
disappearance of the gap in the characteristic Young tableau for the
$U(N)$ representations of angular matrix variables
\ref\buka{D.V.Boulatov and V.A.Kazakov, preprint LPTENS-91/24 (1991),
Int.J.Mod.Phys. A8(1993) 809}.

We can make contact with the results of Gross and Taylor
by expanding the answer about $g^2 A=\infty$.
Although this is a singular point, the form of the singularity allows a
well-defined double expansion in $e^{-A/2}$ and $A e^{-A/2}$, as will emerge in
the following.
{}From \gr, a reason to think that the expansion is unambiguous is that each
term $e^{-nA/2}$ has a coefficient polynomial in $A$ and of order $2n$, a
property we would certainly lose if we expanded the exponentials in some other
way.
A better argument requires knowing the analytic structure of $F(A)$, to which
we turn.
The strong coupling limit was
$\tau\rightarrow 0$; since series expansions of the theta constants are in
$e^{i\pi\tau}$, clearly we want to make a modular transformation.  This can be
done by taking $K\leftrightarrow K'$ and $k\leftrightarrow k'$
and then going to theta functions with modulus $\tau\rightarrow i\infty$ in
the strong coupling limit.
Using Legendre's relation for $E'$ gives
\eqn\areatwo{\eqalign{
{1\over 4}A &= -k^2 K'^2 + 2 K'/K (\pi/2 + K K' - E K')\hfill\cr
&= -\pi i\tau + {(\pi i\tau)^2\over 12}
(\thetaf{3}^4(0|\tau) + \thetaf{4}^4(0|\tau) -2E_2(\tau))\hfill\cr
&= -\pi i\tau -
2\pi i\tau^2 {\partial\over\partial \tau}
\log \thetaf{4}(0|2 \tau)\hfill\cr
&\equiv -\pi i\tau +
(2\pi i\tau)^2 R(2\pi i \tau)}}
Now
\eqn\rseries{\eqalign{R(2\pi i\tau) &= 2 \sum_{n\ge 1} {n e^{2n i\pi\tau}\over
1-e^{4ni\pi\tau}}\hfill\cr
&= 2 e^{2 i\pi\tau} + 4 e^{4 i\pi\tau} + 8 e^{6 i\pi\tau} + 8 e^{8 i\pi\tau} +
12 e^{10 i\pi\tau} + \ldots,}}
so in the limit $A\rightarrow\infty$ we have $\tau=i A/4\pi$ with corrections
exponentially small in $A$.
To get a double expansion in
$\exp -A/2$ and $A \exp -A/2$, we express everything in terms of theta
functions with modulus $2\tau$, and solve for $2\pi i\tau$ in terms of $A$ and
the exponentially small $R$:
\eqn\rsol{\eqalign{
2\pi i\tau &= {1\over 4R}(1-\sqrt{1+4RA})\hfill\cr
&\equiv -\half A~ s(A R)\hfill\cr
&= -\half A(1 - R A + 2R^2 A^2 - 5 R^3 A^3 + 14 R^4 A^4 + \ldots)}}
We would then successively substitute $A$ for $\tau$ in $R$.

Rewriting $F'(A)$ in the same way, we find
\eqn\fprimetwo{\eqalign{
F'(A) &=- {1\over 24} +{1\over 24} s(AR)^2 (\theta_3^4(0|\tau) -
{1\over 2} \thetaf{2}^4(0|\tau)) +
{1\over 1536} s(AR)^4 \thetaf{2}^8(0|\tau) A\hfill\cr
&= - {1\over 24} + {1\over 48} s(AR)^2
(\thetaf{3}^4(0|\tau) + \thetaf{4}^4(0|\tau)) +
{1\over 1536} s(AR)^4 \thetaf{2}^8(0|\tau) A\hfill\cr
&= - {1\over 24}- s(AR)^2 {1\over 4\pi i} {d\over d\tau}
\log {\thetaf{4}(0|2\tau)\over \eta(2\tau)} +
{1\over 96} s(AR)^4 \thetaf{2}^4(0|2\tau) \thetaf{3}^4(0|2\tau) A\hfill\cr
&\equiv - {1\over 24}+s(AR)^2 F'_0(2\pi i\tau) + A~ s(AR)^4 F'_1(2\pi i\tau).}}

We see that the analytic structure of $F'$ in terms of $w=e^{-A/2}$ is not so
simple; however the branch cut in \rsol\ is away from the origin, and near the
origin we have a sum of terms $(\log w)^m f_m(w)$ with each $f_m$ analytic.  If
we were only given the function $F(w)$, we could isolate these terms by
combining its values on the sheets $F(e^{2\pi ik}w)$; at each order in $w$ only
finitely many $f_m$ contribute.  This would fix the double expansion uniquely.

Using \rsol\ and \fprimetwo, we can get the terms at a given order $A^m$ in the
double expansion to all orders in $e^{-A/2}$ by taking $A R$ small;
then (here prime is always $d/dA=-4q~ d/dq$)
\eqn\fseries{\eqalign{
F'(A) = & - {1\over 24}+F'_0(-A/2)
+ A (F'_1 - 2 R F'_0)\hfill\cr
&+ A^2 \left(-4RF'_1+5R^2 F'_0- RF''_0
\right)\hfill\cr
&+ O(A^3)\hfill\cr
= & \sum_{n\ge 1} {(2n-1)e^{-(2n-1)A/2}\over 1-e^{-(2n-1)A/2}}\hfill\cr
&+ {A}\biggl(
{1\over 6}\bigl({1}+8\sum_{n\ge 1}{n e^{-nA/2}(1-(-e^{-A/2})^n)\over 1-e^{-nA}}
\bigr)
\sum_{m\ge 1}{(2m-1)e^{-(2m-1)A/2}\over 1-e^{-(2m-1)A}}
\hfill\cr
&\qquad~~-4\sum_{m\ge 1}{me^{-mA/2}\over 1-e^{-mA}}
\bigl({1\over 24}+\sum_{n\ge 1}{(2n-1) e^{-(2n-1)A/2}\over 1-e^{-(2n-1)A/2}}
\bigr)
\biggr)\hfill\cr
&+ O(A^2).}}
We have compared this with a direct expansion of the formula \parti~ to
$O(\exp(-2A))$:
\eqn\gtexp{\eqalign{
F(A) &= 2e^{-A/2} + (-1-2A+{1\over 2}A^2)e^{-A} +
({8\over 3}+4A^2-{8\over 3}A^3+{1\over 3}A^4)e^{-3A/2} +\ldots\hfill\cr
-F'(A) &= e^{-A/2} + (1-3A+{1\over 2}A^2)e^{-A} +
(4-8A+2A^2-{8\over 3}A^3+{1\over 2}A^4)e^{-3A/2}
+ \ldots,}}
and with the $O(A^0)$ and $O(A)$ terms to much higher order, and found
complete agreement.
One can also reproduce this result by the direct expansion of \fprime~.
\footnote{$\dagger$}{We thank J.-M.Daul,
who reproduced this expansion starting from
the integral  representation
\denfin~ up to $O(\exp(-{3 \over 2}A))$ and found complete agreement.}

We also checked this expansion  from Gross and Taylor's rules,
dropping the terms
involving ``tubes'' and ``handles'' and proportional to powers of $A$ (as
appropriate for $U(N)$).

One technical conclusion we can draw from the solution is that $e^{-A}$ and $A$
are not really the natural expansion parameters in the problem.
One way to think of this is in terms of the formalism of \mrd.  The
perturbation is a higher derivative operator, and it is
perhaps surprising that this even has a non-zero radius of convergence.
Evidently it does, and much of the effect of the perturbation can be expressed
as a ``renormalization'' of the modulus of the cylinder from $A$ to the
variable $\tau$ determined as above.

In Gross and Taylor's language the power-like expansion in $A$ is given by
inserting branch points on the string world-sheet.  Evidently this expansion
diverges at $A_c$.

We might hope to see the transition from the weak coupling side by noticing
that some expectation value obtains an impossible value in the continuation
beyond $A_c$, just as for the GWW transition, the continuation of the strong
coupling spectral density to small $g$ was no longer positive.
\ref\Friedan{D. Friedan, Comm. Math. Phys. 78, 353-362 (1981).}
The easy observables to compute here are expectation values of the local
operators $\tr E^k$ ($E$ is the electric field), given by $\int dh u(h) h^k$.
Another possible way to see the transition would be to see the sum of terms
$e^{-N/A}$ in \groupasy\ diverge for sufficiently large $A$.

Another quantity revealing of the weak coupling phase is the Wilson loop,
separating regions of area $A_1$ and $A_2$.
This is again simple in terms of the heat kernel:
\eqn\wils{W_{G=0}(A_1,A_2) = \int dU~G(1,~U;\ A_1/2N)~~\half\tr (U+U^{+})~~
G(U,~1;\ A_2/2N).}
{}From \groupasy\ we learned that in the weak coupling phase, only the leading
classical solution contributes in the heat kernel.  Thus we can take the
expression of \meno\ and drop the winding terms:
\eqn\green{G(1,U;A/2N) = \CN \prod_{i<j}
{\theta_i-\theta_j\over \sin \half(\theta_i-\theta_j)}
e^{-(N/A) \sum_i \theta_i^2}}
where $e^{i\theta_i}$ are the eigenvalues of $U$.
Since the invariant measure is
\eqn\meas{\int dU = \int \prod
d\theta_i~\prod_{i<j}\sin^2\half(\theta_i-\theta_j),}
\wils\ becomes
\eqn\wint{W_{G=0}(A_1,A_2) = \int \prod d\theta_i~
\prod_{i<j} (\theta_i-\theta_j)^2 e^{-N({1\over A_1}+{1\over A_2})\sum_i
\theta_i^2} \sum_i\cos \theta_i}
which is again an expectation value at a semicircular saddle point:
\eqn\wexp{\eqalign{
W_{G=0}(A_1,A_2) &= \left({2\over\sqrt{x}}\right) J_1(\sqrt{x})\hfill\cr
&= 1 - x/8 + \ldots}}
where $x = A_1 A_2 / (A_1+A_2)$.

This is perhaps a peculiar result.  It is positive for all $A_1+A_2<A_{crit}$
but becomes oscillatory for (unphysical) large $A$.  Even stranger, its
asymptotic behavior is $\cos(\sqrt{A})/A$.

It would be very interesting to calculate the Wilson loop in the strong
coupling phase.
Knowing the result for the simple loop,
the two-dimensional renormalized loop equations of
\kkone~ would then determine all loop averages.
It is clear that these will obey the same Gross and Taylor
rules of the large $A_1,A_2$ expansion (but the coverings will have the
topology of the disc now). In the limit of large total area of the
sphere we would reproduce all the results of \kkone~ for the Wilson
average on the infinite plane. Only the coverings which do not wind
over the sphere will survive, and they are precisely those which were
observed in \kkone~, \koone~.

Returning to the general consequences of the phase transition,
the conclusion for the relation of the strong coupling expansion to string
theory is that string rules derived from the heat kernel action by expanding
about $g^2=\infty$ (in \gt, the expansion was in $\exp -g^2$), do not give the
correct answer on a small two-sphere.
Now if we were two-dimensional, we might not care about this -- the string
rules DO give the right answer in the $A\rightarrow\infty$ limit, and all we
seem to need for this is that the overall area of our universe be large.
Expectation values for a Wilson loop enclosing an arbitrary area in this
universe will be given correctly.
Certainly the 't Hooft model of mesons in QCD$_2$ has no large $N$ transition
in infinite Minkowski space.


For four dimensional physicists, however, this transition looks like a real
problem.  The only precise way we know to define the strong coupling expansion
is to start on a lattice, and take the continuum limit as defined by Wilson.
For QCD this will require taking the bare coupling to zero in the way
prescribed by the RG, so our expansion must make sense at weak coupling.
We therefore need to choose an action for which there is (among other
constraints) no large $N$ transition.
Although we do not know if the transition we find for the heat kernel action
persists in $D>2$, the simple fact that any higher dimensional lattice contains
embedded topological two-spheres puts the burden of proof on the other side --
to show that somehow cancellations between terms in the higher dimensional
series eliminate the transition.

\vskip 0.3in
We thank T.Banks, E.Brezin, D.Boulatov,
J.-M.Daul, J.Distler, P.Ginsparg, D.Gross, C.Itzykson, I.Kostov,
H.Neuberger, S.Shenker and M.Staudacher for enjoyable discussions.

\vfill\eject
\appendix{A}{Elliptic Integrals and Theta Functions}
The basic integrals required are
\eqn\intzero{I_0 = \int_{-b}^b
{d\lambda\over\sqrt{(a^2-\lambda^2)(b^2-\lambda^2)}} = {2\over a}K(b/a)}
\eqn\inttwo{I_2 = \int_{-b}^b
{d\lambda ~\lambda^2\over\sqrt{(a^2-\lambda^2)(b^2-\lambda^2)}} =
2a[K(b/a)-E(b/a)]}
and
\eqn\intfour{I_4 = \int_{-b}^b
{d\lambda ~\lambda^4\over\sqrt{(a^2-\lambda^2)(b^2-\lambda^2)}} = {2\over
3}a[(2a^2+b^2)K(b/a)-2(a^2+b^2)E(b/a)]}
in terms of the standard complete elliptic integrals with modulus $k=b/a$
(e.g. as in \ref\bateman{A. Erd\'elyi et. al., Higher Trancendental Functions,
McGraw-Hill, 1953.})

We will write the results in terms of the complementary modulus $k'$ satisfying
$k'^2=1-k^2$, using $K(k)=K'(k')$, $E(k)=E'(k')$ and re-express them as theta
constants, for the strong coupling expansion.  From \bateman~ (13.20) we have
\eqn\thetadef{\eqalign{
k &= \thetaf{2}(0|\tau)^2/\thetaf{3}(0|\tau)^2\hfill\cr
k' &= \thetaf{4}(0|\tau)^2/\thetaf{3}(0|\tau)^2\hfill\cr
K(k) &= {\pi\over 2}\thetaf{3}^2(0|\tau)\hfill\cr
K'(k) &= {-i\pi\tau\over 2}\thetaf{3}^2(0|\tau)\hfill\cr
E(k) &= {\thetaf{3}^4(0|\tau) + \thetaf{4}^4(0|\tau)\over 3
\thetaf{3}^4(0|\tau)} K(k) - {1\over 12 K(k)}
{\thetaf{1}'''(0|\tau)\over\thetaf{1}'(0|\tau)}}}
where $q = e^{i\pi\tau} = \exp(-\pi K'/K)$.

Some standard identities involving these functions:
\eqn\thetaident{\eqalign{
&K E' + K' E - K K' = \half\pi\hfill\cr
&{\partial^2\over\partial\nu^2}\theta_n =
4\pi i{\partial\over\partial\tau}\theta_n \qquad \forall n\hfill\cr
&\thetaf{1}'(0|\tau) = \pi
\thetaf{2}(0|\tau)\thetaf{3}(0|\tau)\thetaf{4}(0|\tau)
= 2\pi \eta(\tau)^3\hfill\cr
& \thetaf{3}^4(0|\tau)=\thetaf{2}^4(0|\tau) + \thetaf{4}^4(0|\tau)\hfill\cr
&{\thetaf{1}'''(0|\tau)\over\thetaf{1}'(0|\tau)} =
4\pi i{\partial\over\partial\tau} \log \thetaf{1}'(0|\tau) =
12\pi i{\partial\over\partial\tau} \log \eta(\tau) = -\pi^2 E_2(\tau)\hfill\cr
}}
where $E_2$ is the normalized Eisenstein series.
We will need as well
\eqn\thetaidt{\eqalign{
\thetaf{3}^4(0|\tau) &= {4\over \pi i}{\partial\over\partial\tau}
\log {\thetaf{2}(0|\tau)\over\thetaf{4}(0|\tau)}\hfill\cr
\thetaf{2}^4(0|\tau) &= {4\over \pi i}{\partial\over\partial\tau}
\log {\thetaf{3}(0|\tau)\over\thetaf{4}(0|\tau)}}}
proven by checking that both sides are modular forms of weight $2$ and level
$2$, and low orders in the $q$-expansion.\hfill\break
All of the relevant theta constants can be expressed in terms of ones with
modulus $2\tau$:
\eqn\thetatwo{\eqalign{
\thetaf{3}^4(0|\tau) &=
(\thetaf{3}^2(0|2\tau)+\thetaf{2}^2(0|2\tau))^2\hfill\cr
\thetaf{4}^4(0|\tau) &=
(\thetaf{3}^2(0|2\tau)-\thetaf{2}^2(0|2\tau))^2\hfill\cr
\thetaf{3}^4(0|\tau) &+ \thetaf{4}^4(0|\tau)
= {8\over \pi i}{\partial\over\partial\tau}
\log
{\thetaf{2}(0|2\tau)\thetaf{3}(0|2\tau)\over\thetaf{4}^2(0|2\tau)}\hfill\cr
&\qquad\qquad\    = -{12\over \pi i} {\partial\over\partial\tau} \log
{\thetaf{4}(0|2\tau)\over\eta(2\tau)}\hfill\cr
E_2(\tau) &= {6\over \pi i} {\partial\over\partial\tau} \log
{\thetaf{4}(0|2\tau)\eta(2\tau)}}}
The first two follow from \bateman\ 13.23.15;
the third uses these, \thetaidt\ and Jacobi's identity (line 3 in \thetaident);
the fourth follows from substituting the product representations below.

We will then need series expansions of these and their $\tau$-derivatives.
These are best derived from the logarithmic derivatives.  (Here $q =
e^{i\pi\tau}$):

\eqn\thetaseries{\eqalign{
-{1\over 4\pi i}{\partial\over\partial\tau} \log
{\thetaf{4}(0|2\tau)\over\eta(2\tau)} &=
{1\over 24}+\sum_{n\ge 1} {(2n-1)q^{4n-2}\over 1-q^{4n-2}}\hfill\cr
-{1\over 4\pi i}{\partial\over\partial\tau} \log \thetaf{4}(0|2\tau) &=
\sum_{n\ge 1} {n q^{2n}\over 1-q^{4n}}\hfill\cr
\thetaf{2}^4(0|2\tau)
&= 16 \sum_{n\ge 1}{(2n-1) q^{4n-2}\over 1-q^{8n-4}} \hfill\cr
\thetaf{3}^4(0|2\tau) &= 1+
8 \sum_{m\ge 1} {m q^{2m}(1-(-q^2)^{m})\over 1-q^{4m}}\hfill\cr
E_2(2\tau) &= 1 - 24 \sum_{m\ge 1} {m q^{4m}\over 1-q^{4m}}.}}

Finally, we use product representations like
\eqn\thetapro{\eqalign{
\theta_4(0|\tau) &= \prod_{n\ge 1} (1-q^{2n})(1-q^{2n-1})^2\hfill\cr
\eta(\tau) &= q^{1/12} \prod_{n\ge 1} (1-q^{2n}).}}

\listrefs
\end